\documentclass{jfm}

\usepackage{graphicx}
\usepackage{epstopdf, epsfig}
\usepackage{newtxtext}
\usepackage{newtxmath}
\usepackage{natbib}
\usepackage{hyperref}
\usepackage{siunitx}
\usepackage{wasysym}
\usepackage{ulem} 
\usepackage{bm} 
\usepackage{xcolor}
\DeclareSIUnit \pixel {px}
\hypersetup{
    colorlinks = true,
    urlcolor   = blue,
    citecolor  = blue,
}
\shorttitle{Spreading drops of granular suspensions}
\shortauthor{A. Pelosse, \'E. Guazzelli and M. Roch\'e}

\title{Probing dissipation in spreading drops with granular suspensions}

\author{Alice Pelosse, \'Elisabeth Guazzelli,
\and Matthieu Roch\'e\corresp{\email{matthieu.roche@u-paris.fr}}}

\affiliation{Universit\'{e} Paris Cit\'{e}, CNRS,  Mati\`{e}re et Syst\`{e}mes Complexes UMR 7057, F-75013, Paris, France}

\begin{document}

\maketitle

\begin{abstract}
In this article, we study the spreading of droplets of density-matched granular suspensions on the surface of a solid. Bidispersity of the particle size distribution enriches the conclusions drawn from monodisperse experiments by highlighting key elements of the wetting dynamics. In all cases, the relation between the dynamic contact angle and the velocity of the contact line follows a similar relation as that of a simple fluid, despite the complexity introduced by the presence of particles. We extract from this relation an apparent wetting viscosity of the suspensions that differs from that measured in the bulk. Dimensional analysis supported by experimental measurements yields an estimate of the size of the region inside the droplet where the value of the dynamic contact angle depends on a balance of viscous dissipation and capillary stresses. Depending on how particle size compares with this viscous cut-off length seems crucial in determining the value of the apparent wetting viscosity. With bimodal blends, the particle size ratio can be used to show the effects of the local structure and volume fraction at the contact line, both impacting the value of the corresponding wetting viscosity.
\end{abstract}

\section{Introduction}
\label{sec:intro}

A liquid wetting the surface of a solid is a common observation in daily experiences. A classical example is that of raindrops hitting a glass window and falling along its surface, leaving trails of water. The physics underlying this supposedly simple situation is in fact quite rich as the properties of all the involved media matter, including those of the surrounding atmosphere. Moreover, length scales from the molecular up to the millimetre range must be considered, leading to a high level of theoretical complexity. A drop of simple fluid spreading onto a solid substrate is subjected to three competing physical mechanisms, namely gravity, capillarity, and viscosity \citep[see e.g.\,][]{bonn2009wetting}. The shape and dynamics of the droplet depend on the length scale at which they are analysed with respect to the capillary length, $\ell_c=(\gamma/\rho g)^{1/2}$, with $\gamma$ the surface tension of the solid, $\rho$ its density, and $g$ the acceleration of gravity. If we consider droplets with a radius $R>>\ell_c$, the balance between gravity and capillarity controls the shape and dynamics at length scales comparable to $R$. At microscopic scales near the triple-phase contact line, the balance involves instead viscosity and capillarity. As the height diminishes at the approach of the contact line, viscous stresses diverge when using the classical no-slip boundary condition on the substrate. A variety of models have been proposed to regularise this non-integrable singularity. A common way to circumvent this divergence issue has been to replace the no-slip boundary condition by a Navier slip condition, i.e. to introduce a microscopic cut-off length scale, $\lambda$, in the continuum modelling \citep{huh1971hydrodynamic,dussanv1974}. For the intermediate scales, gravity and viscosity as well as capillarity participate into the dynamics of the drop. 

The above picture becomes more involved when the spreading fluid is complex, such as a polymer solution \citep{lee2022effect}, a polymeric melt \citep{seemann2005dynamics}, or a viscoplastic material \citep{spaid1996stability}. Such fluids can have intrinsic timescales (e.g.\,relaxation, agitation) or characteristic length scales (e.g.\,particle size, persistence length) that need to be accounted for in the spreading dynamics. The present work focuses on the spreading of dense granular suspensions of large spherical particles that do not experience Brownian motion. These particulate systems are characterised by an additional length scale, the particle size, that must be addressed in the multiscale description of an advancing contact line. The interplay of the vanishing height of the flow and the finite size of the particles should control the distance at which the particles can approach the contact line. This view has been confirmed in our previous study of the motion of the triple-phase contact line surrounding a droplet of monomodal granular suspension \citep{zhao2020spreading}. Interestingly, the relation between the dynamic contact angle, i.e.\, the angle between the liquid-gas interface and the solid-liquid interface measured in the droplet, and the speed of the contact line happens to be similar to that found for a simple liquid, known as the Cox-Voinov law \citep{voinov1976hydrodynamics,cox1986dynamics}. In dimensionless form, this speed was reported in terms of the capillary number, $Ca=\eta U/\gamma$, measuring the relative effect of viscous to surface tension forces,  where $\eta$ is the viscosity of the liquid and $U$ the velocity of the contact line. We found that the viscosity involved in this capillary number differed from the widely-studied bulk viscosity of suspensions as it depended on particle diameter, $d$, in addition to particle volume fraction, $\phi$. This observation resulted from the aforementioned ability of the particles to approach the contact line close enough to affect dissipation. In particular, we showed that the apparent viscosity reduced to the viscosity of the suspending fluid when the particle size became larger than approximately 100 \si{\micro\meter}. This study suggests that a granular suspension may be an interesting system for the study of wetting as the discrete nature of this complex fluid can be used to probe the size of the domain in which the Cox-Voinov relation is a valid description of the relation between the dynamic contact angle and the velocity of the contact line.

Another feature specific to the spreading of a granular suspension is that there is a self-organisation of the advancing front rows of particles near the rim of the drop  \citep{zhao2020spreading}. To give a full picture, the close vicinity of the contact line is a region devoid of particles.  Behind this particle-depleted region, a few layers of crystallised beads are observed.
As the height increases further, the particles switch from a crystal-like to a disordered structure. This ordering is particularly seen for dense suspensions, e.g.\, for packing fractions, $\phi$, of 40\% or above. Confinement by the free interface seems to be responsible for this observed organisation. However, the role of this ordered particle phase in energy dissipation and its effects on the dynamics of spreading are still open problems that require scrutiny. A simple model that matches the shape of the droplet in the depleted region to that of a particle-rich region with no constraint on the ordering of particles fails at describing the relation between contact angle and velocity \citep{zhao2020spreading}.

Particles near a receding contact line have been experimentally studied when a plate is withdrawn from a bath of non-Brownian suspension. In this configuration named dip-coating, the key result is that the plate can be coated by particles when the entrained film is thicker than the particle diameter. More precisely, increasing the withdrawal speed leads to different regimes of coating going from a film devoid of particles to a heterogeneous monolayer of particles, and finally to a thick film of suspension entrained by the plate \citep{gans2019dip,palma2019dip}. When the suspension contains two sizes of particles, the small particles are first entrained on the plate to form a heterogeneous monolayer and the large particles begin to be entrained only with a further increase in withdrawal speed \citep{Jeong_dip_2022}. In the heterogeneous regime, confined particles gather in clusters without clear ordering in contrast to the highly-structured phase observed near the advancing contact line during the spreading of suspension drops on a solid substrate \citep{zhao2020spreading}.

In this work, we examine the advancing contact line of a dense granular suspension, with the aim of clarifying the origin of the difference between bulk viscosity and its counterpart extracted from wetting experiments. In \S\,\ref{sec:theory}, we discuss the range of interface heights for which we should expect the viscous-capillary balance at play in the Cox-Voinov relation to be valid. The materials and methods used in the experiments are described in \S\,\ref{sec:exp}.  In \S\,\ref{sec:wetting}, we test experimentally the predictions of \S\,\ref{sec:theory} with simple fluids. Then, we resort to the discrete nature of the particles in a suspension to probe dissipation in the vicinity of the contact line. These experiments are another test of the outcomes of \S\,\ref{sec:theory}. Results obtained for monodisperse and bidisperse suspensions consisting of different particle combinations are presented in \S\,\ref{sec:susp_wetting}. While bidisperse suspensions seem to be a more complicated system, they offer the possibility to use two different sizes for scrutinising dissipation. We discuss our findings and provide concluding remarks in \S\,\ref{sec:conclusion}.
\section{Elements of the wetting theory}
\label{sec:theory}
%
\subsection{General framework}
\label{subsec:framework}
\begin{figure}
    \centering
    \includegraphics[width=0.8\linewidth]{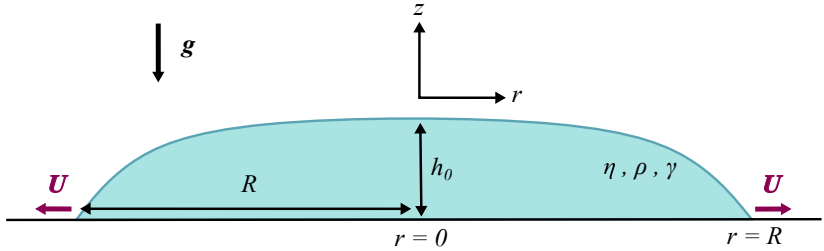}
    \caption{Sketch of a fluid droplet spreading on a rigid substrate.}
    \label{fig:dropletSchematic}
\end{figure}

 The classical situation of a fluid droplet of radius $R$ and central height $h_{0}$ that spreads onto the surface of a rigid substrate is depicted in figure \ref{fig:dropletSchematic}. The fluid has a dynamic viscosity $\eta$, a density $\rho$, and a surface tension $\gamma$. In most situations, inertia is negligible, i.e. the Reynolds number is small, ${Re}=\rho Uh_0/\eta\ll 1$, where $U$ is the characteristic spreading velocity. Since $h_0\ll R$, the lubrication approximation can be applied. As the flow is axisymmetric,  we use cylindrical coordinates with the $z$-axis being the axis of symmetry. Under these assumptions, the height of the air-liquid interface, $h(r)$, is a solution of the free-surface, thin-film equation,
\begin{equation}
    \frac{3\eta}{\gamma}\partial_{t}h+\frac{1}{r}\partial_{r}\left[h^2(h+3\lambda)r\partial_{r}\left(\partial_{rr}h+\frac{1}{r}\partial_{r}h-\frac{\rho g}{\gamma}h \right) \right]=0,
    \label{eq:dynamicDroplet}
\end{equation}
where $g$ is the gravitational acceleration \citep{hocking1981,hocking1983,savva2012}. A particular feature of the above equation \eqref{eq:dynamicDroplet} is that a characteristic length scale, $\lambda$, has been introduced to relax the no-slip boundary condition at the line of contact between the droplet, the substrate, and the surrounding gas \citep{huh1971hydrodynamic,dussanv1974}. In the context of the present work, we grant $\lambda$ its usual role of a slip length and we take it to be of the order of the size of a few nanometers, comparable to that of a molecule of the liquid. 

The resolution of \eqref{eq:dynamicDroplet} requires appropriate boundary conditions to be specified as
\begin{equation}
    h(r\rightarrow R)=0,\; \partial_rh(r\rightarrow R)=\theta_m,\; \mbox{and} \; \int_{r=0}^{R}h(r)2\pi rdr={V}_0,
    \label{eq:boundary_conditions}
\end{equation}
where $V_0$ is the volume of the droplet and $\theta_m$ is the microscopic contact angle.

Rather than providing  detailed derivations of the solutions of \eqref{eq:dynamicDroplet} with these boundary conditions \eqref{eq:boundary_conditions} \citep[see e.g.][]{hocking1983,cox1986dynamics,sibley2015asymptotics,savva2009two}, we give some insight into the different length scales involved in the problem described by \eqref{eq:dynamicDroplet} with \eqref{eq:boundary_conditions}. Besides the slip length, $\lambda$, we also identify the capillary length, $\ell_{c}=(\gamma/\rho g)^{1/2}$, for which gravitational forces balance capillary forces, and the radius of a sphere having the same volume ${V}_0$ as that of the droplet, $R_0=(3{V}_0/4\pi)^{1/3}$. In experiments, it usually happens that $\ell_{c}\gg\lambda$ and $R_0\gg\lambda$ by several orders of magnitude. There is thus a region in the close vicinity of the contact line in which the force balance only involves viscous and capillary forces, with slip being dominant and gravity negligible. This viscous-capillary region will be introduced in \S\,\ref{subsec:inner}. On the opposite range, the large scales are governed by the balance between gravity and capillarity. This capillary-gravity region will be presented in \S\,\ref{subsec:outer}. In between, the three contributions of the competing viscous, capillary, and gravity forces must be retained as developed in \S\,\ref{subsec:intermediate}.
\subsection{Viscous-capillary region}
\label{subsec:inner}

This region corresponds to the very close vicinity of the contact line, i.e. at length scales much smaller than $\ell_c$. As mentioned earlier in \S\,\ref{subsec:framework}, the introduction of a Navier slip is an important ingredient as viscous dissipation would diverge as $h\rightarrow 0$ otherwise. Resolution of the reduced equation \eqref{eq:dynamicDroplet} where the gravity term has been dropped, with the additional assumption that spreading is quasi-steady, leads to a relation often referred to as the Cox-Voinov law \citep{voinov1976hydrodynamics,cox1986dynamics,snoeijer2006b,bonn2009wetting} between the velocity, $U$, of the contact line and the dynamic contact angle, $\theta_{app}(x)$, measured at a distance $x=R-r$ from the moving rim of the droplet,
\begin{equation}
    \theta_{app}^3(x)=\theta_{m}^3+9 {Ca}\log{\left(\frac{x}{\lambda}\right)},
    \label{eq:voinovTannerLaw}
\end{equation}
where $Ca=\eta U/\gamma$ is the capillary number. In the case of viscous liquids, the microscopic contact angle, $\theta_m$, is close to the static value \citep{bonn2009wetting}. For (nearly) perfectly wetting liquids, it is found that  $\theta_m\ll1$ is therefore negligible in \eqref{eq:voinovTannerLaw}. Equation \eqref{eq:voinovTannerLaw} indicates that $\theta_{app}(x)$ is an increasing function of the distance $x$ to the contact line at a given capillary number, ${Ca}$. The interface in this viscous-capillary region must then have a positive curvature when measured in the frame defined in figure\,\ref{fig:dropletSchematic}.
\subsection{Capillary-gravity region}
\label{subsec:outer}

The opposite limit corresponds to the region where $r\rightarrow 0$, i.e.\,$x\sim R\sim R_0$. Moreover, we are interested in the limit where $R_0>>\ell_c$.
At these scales, viscous forces are insignificant and the time-derivative term in \eqref{eq:dynamicDroplet} can be dropped. 
In addition, in this macroscopic region, $\lambda$ can be neglected in front of $h$ which is of the order of $h_0$. Hence, we are left with a capillary-gravity balance, and \eqref{eq:dynamicDroplet} reduces to
\begin{equation}
    \partial_{r}\left[h^3r\partial_{r}\left(\partial_{rr}h+\frac{1}{r}\partial_{r}h-\frac{\rho g}{\gamma}h \right) \right]=0.
    \label{eq:staticDroplet}
\end{equation}
This equation can be integrated once. Using the boundary condition $h(r=R)=0$, the integration constant is seen to be zero. At that stage of the calculation, it is convenient to normalise the $r$- and $h$-scales as $r=R_0\tilde{r}$ and $h=h_{0}\tilde{h}$ where $R_0$ and $h_0$ are the radius of the spherical drop and the characteristic interface height in this region, respectively. Performing another integration leads to
\begin{equation}
    \partial_{\tilde{r}\tilde{r}}\tilde{h}+\frac{1}{\tilde{r}}\partial_{\tilde{r}}\tilde{h}-{Bo}\tilde{h} =C,
    \label{eq:nondimStaticDroplet}
\end{equation}
where $C$ is a constant and $Bo=\rho g R_0^2/\gamma=\left(R_0/\ell_c\right)^2$ the Bond number of the droplet.
The exact solution of \eqref{eq:nondimStaticDroplet} with the boundary conditions \eqref{eq:boundary_conditions} is
\begin{equation}
    \tilde{h}(\tilde{r})\propto\frac{I_0\left(Bo^{1/2}\frac{R}{R_0}\right)-I_0\left(Bo^{1/2}\tilde{r}\right)}{I_2\left(Bo^{1/2}\frac{R}{R_0}\right)},
    \label{eq:sol_static_adim}
\end{equation}
where $I_n(x)$ is the $n$-th modified Bessel function of the first kind \citep{hocking1983}. The solution \eqref{eq:sol_static_adim} reads with dimensional coordinates
\begin{equation}
    h(r)=\frac{{V}_0}{\pi R^2}\frac{I_0\left(\frac{R}{\ell_c}\right)-I_0\left(\frac{r}{\ell_c}\right)}{I_2\left(\frac{R}{\ell_c}\right)}.
    \label{eq:sol_static}
\end{equation}
Droplets are spherical caps when $Bo\ll 1$ whereas their top surface flattens and they look like puddles in the limit ${Bo}\gg 1$. In the frame of reference defined in figure \ref{fig:dropletSchematic},  the shape of the interface in the macroscopic region has a negative curvature.
\subsection{Viscous-capillary-gravity region}
\label{subsec:intermediate}

We finally turn to the examination of the region for which all the contributions of the viscous, capillary, and gravity forces must be kept in \eqref{eq:dynamicDroplet}. In this viscous-capillary-gravity region, the length scale is comparable neither to $\lambda$ nor to $R_{0}$ and the sole remaining length scale is the capillary length. Since $x$ and $h\gg\lambda$, the reduced equation then reads
\begin{equation}
    3\,{Ca}\,\partial_{r}h+\frac{1}{r}\partial_{rr}\left[h^3x\partial_{r}\left(\partial_{rr}h+\frac{1}{r}\partial_{r}h-\frac{h}{\ell_{c}^2} \right) \right]=0,
    \label{eq:dynamicDroplet_vgc}
\end{equation}
where we consider quasi-static spreading and use $\partial_{t}h=U\partial_{r}h$. This assumption  means that at any time, the velocity of the contact line sets the drop shape with no transient involved. This constraint is enforced by the hypothesis that $Re\ll1$ \citep{gratton1996}. The  capillary number, $Ca=\eta U/\gamma$, then appears as a natural way to build a dimensionless velocity.

As mentioned above, the capillary length, $\ell_c$, is the only length available. It is the scale for the variation along the horizontal direction, $r=\ell_{c}\hat{r}$.  However, no obvious length scale emerges in the vertical direction. We thus define $h=h^\star\hat{h}$ where $h^\star$ is the (still unknown) relevant scale for the drop height. Using these renormalisations in \eqref{eq:dynamicDroplet_vgc} yields
\begin{equation}
    3Ca\partial_{\hat{r}}\hat{h}+\left(\frac{h^\star}{\ell_{ c}}\right)^3\frac{1}{\hat{r}}\partial_{\hat{r}}\left[\hat{h}^3\hat{r}\partial_{\hat{r}}\left(\partial_{\hat{r}\hat{r}}h+\frac{1}{\hat{r}}\partial_{\hat{r}}\hat{h}-\hat{h} \right) \right]=0.
    \label{eq:nondimVGCdomain}
\end{equation}
For the two terms to be of the same order in \eqref{eq:nondimVGCdomain}, one must take the height scale to be 
\begin{equation}
    h^{\star}\equiv\ell_{c}Ca^{1/3}.
    \label{eq:vgcScale}
\end{equation}

This scale represents the typical height separating the viscous-capillary region governed by the Cox-Voinov law from the viscous-capillary-gravity region where gravity comes into play in the force balance. In other words, this transition scale corresponds to the upper bound of the region of the drop where the Cox-Voinov relation \eqref{eq:voinovTannerLaw} is still valid. Another interesting interpretation of $h^\star$ is that it may delineate the change in surface curvature and thus can be understood as the inflection point of the drop interface. 

Some orders of magnitude can be provided for the present experimental conditions. The range of $h^\star$ is $20\leq h^\star \leq 800$ \si{\micro\meter} for typically $\ell_c\simeq 2$~mm and  $10^{-6}\leq Ca\leq10^{-2}$. As a consequence, liquids with sub-millimetre characteristic length scales, such as suspensions of non-Brownian particles, may show non-trivial $\theta_{app}-Ca$ relations. Indeed, adding density-matched particles should not modify the drop behaviour in the gravity driven-region while it should enhance dissipation provided the particles can access the region where viscosity matters, namely the dissipation region for $h\lesssim h^\star$.

\section{Experimental methods}
\label{sec:exp}
%
\subsection{Particles and fluid}

The suspending fluid is a Newtonian PEG copolymer, Poly(ethylene glycol-ran-propylene glycol) monobutyl ether (average $M_n\simeq 3900$, Sigma-Aldrich reference 438189), with a density close to that of polystyrene, $\rho=1056$~kg/m$^3$, at a temperature of \SI{25}{\celsius}. Its dynamic viscosity, $\eta_f=2.4\pm 0.1$~Pa.s, is constant over a large range of shear rate (0.01-10 s$^{-1}$) at \SI{25}{\celsius}. Particles are spherical polystyrene beads (Dynoseeds TS, Microbeads, Norway) that are sieved when necessary to remove small dust particles or to narrow the size distribution below a tenth of particle diameter. The mean particle diameters, $d$, used in the experiment are 10, 20, 40, 80, 140 and \SI{250}{\micro\meter}. Density matching between the liquid and the particles is sufficient to prevent buoyancy effects over hours at the temperature of \SI{25}{\celsius} in the air-conditioned room. To prepare the suspension mixture, the suspending fluid is weighted and poured into a test tube. A desired mass of particles is then added to reach the target volume fraction, $\phi$. Good mixing  while avoiding entrapping of air is achieved by hand mixing followed by slow mixing on a rolling device overnight. In the following experiments, the solid volume fraction is $\phi=0.4$ as we aim at studying the strongest effects expected at large volume fractions \citep{zhao2020spreading}. The particle surface is completely wet by the suspending fluid, i.e. particles remain suspended in the fluids and do not aggregate. We measured the surface tension of the suspension using pendant drop experiments. We find that the suspension surface tension is equal to the surface tension of the suspending fluid, $\gamma=\gamma_f\simeq$~\SI{35}{\milli\newton\per\meter},  a similar result to that reported in \cite{couturier2011suspensions}.

\subsection{Bulk suspension rheology}

\begin{figure}
    \centering
    \includegraphics[width=0.5\linewidth]{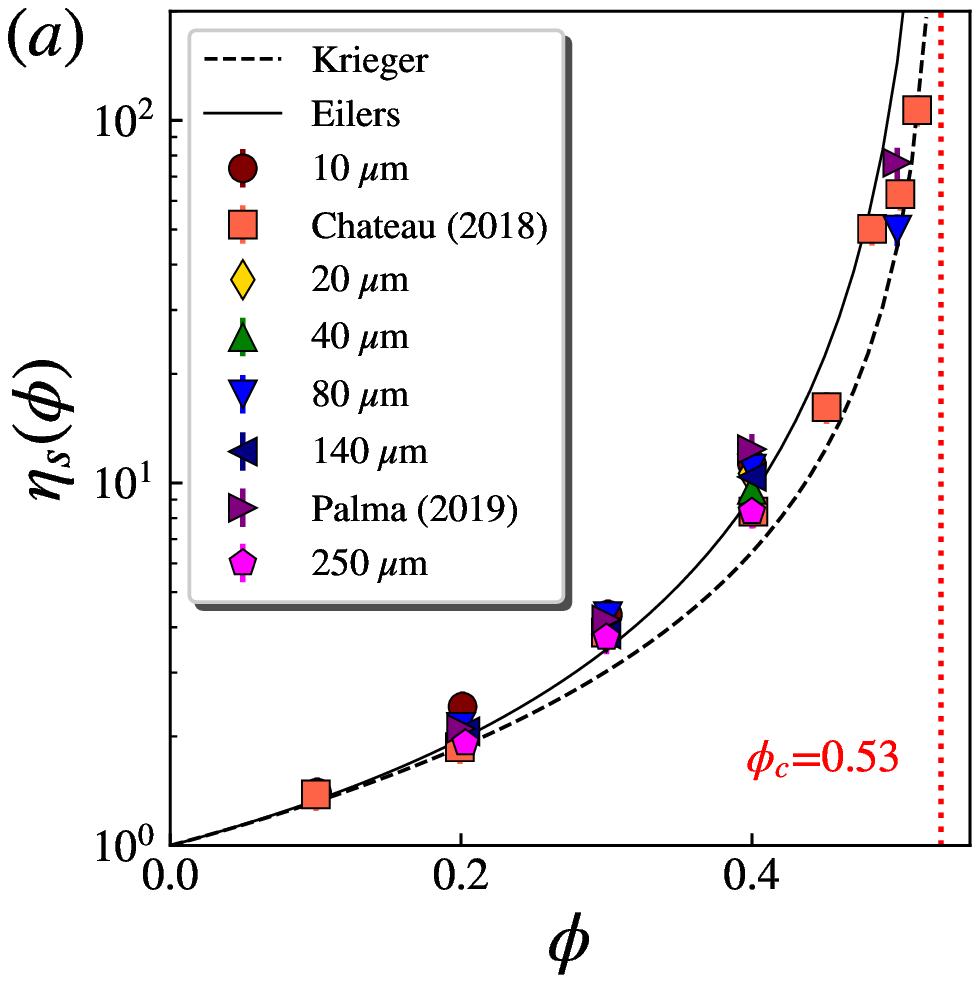}\hfill
    \includegraphics[width=0.5\linewidth]{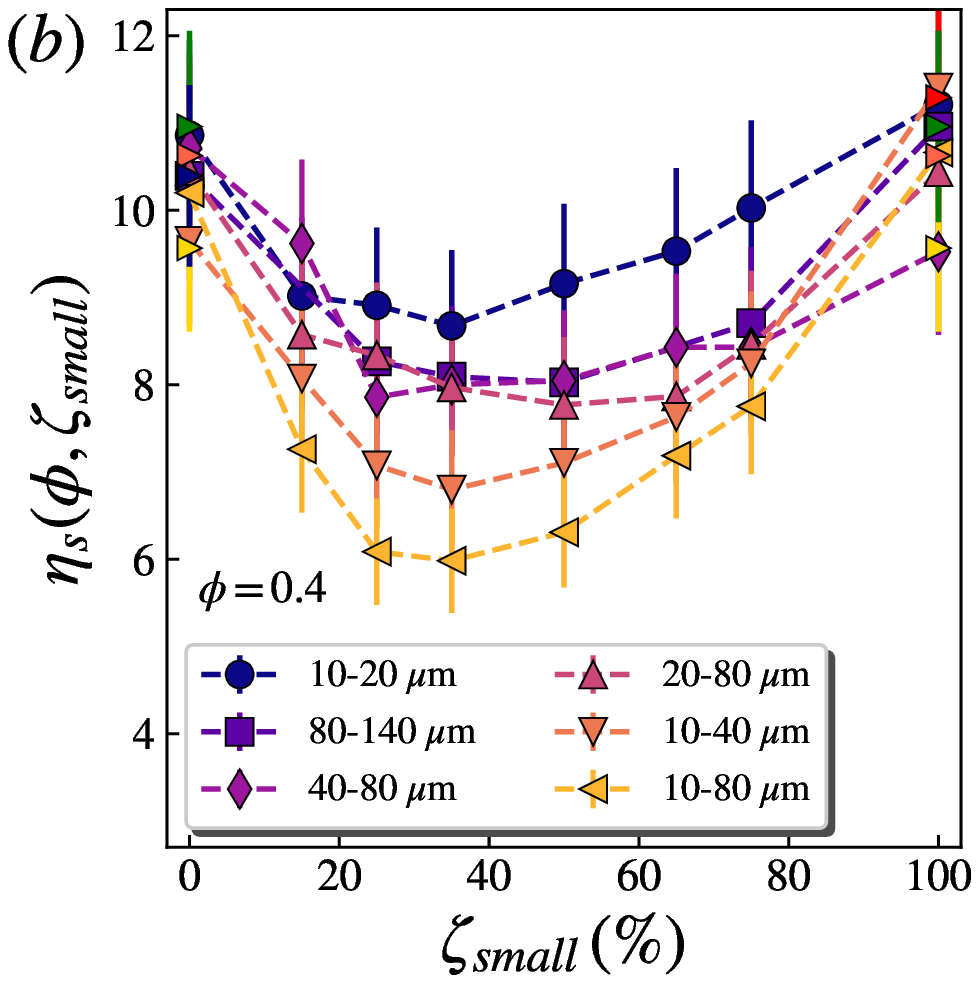}
    \caption{Relative viscosity (i.e. shear viscosity of the suspension relative to that of the suspending fluid), $\eta_s$, of $(a)$ monomodal and $(b)$ bimodal suspensions averaged for shear rates between \SI{0.1}{\per\second} and \SI{1}{\per\second} corresponding to the range of shear rates of the spreading experiments. $(a)$ Relative viscosity of monomodal suspensions as a function of particle volume fraction, $\phi$, for particles with diameters of $10,\,20,\,40,\,80,\,140,$~\SI{250}{\micro\meter} and comparison with experimental data for particles with diameters of $10$ and \SI{140}{\micro\meter} from the experiments of \cite{chateau2018pinch} and \cite{palma2019dip}, respectively. The vertical dotted line is an estimate of the maximum volume fraction, $\phi_c\simeq0.53$, used to compare data with the empirical correlations of Krieger-Dougherty, $\eta_s=(1-\phi/\phi_c)^{-[\eta]\phi_c}$ \citep{krieger1959mechanism}, and Eiler, $\eta_s=(1+[\eta]/2\phi/(1-\phi/\phi_c))^2$  \citep{eilers1941viskositat} ($[\eta]=2.5$ is the intrinsic viscosity of the suspension). $(b)$ Relative viscosity of bimodal suspensions at a fixed total solid volume fraction of $\phi=0.4$ as a function of the fraction of the small particles in the solid phase, $\zeta_{small}$, for two-size blends (see legend).}
    \label{fig:rheology}
\end{figure}

A thorough discussion of our results requires a comparison of the bulk viscosity of the granular suspensions with their apparent viscosity extracted from drop spreading experiments, see \S\ref{sec:susp_wetting}. Rheological measurements have been performed using an ARES G2 rheometer (TA Instruments) with a 25\,mm-wide plate-plate geometry for solid blends of moderate-size particles with diameters of 10, 20, 40, and \SI{80}{\micro\meter}. The thickness of the gap between the plates is typically \SI{1.5}{\milli\meter}, i.e. at least 20 particle diameters, to prevent confinement effects \citep{peyla2011new}. As wall-slip effects depend on the gap thickness, the fact that no viscosity variation is measured while changing the gap thickness from 1~mm to 2~mm indicates that sliding is negligible for these small particles \citep{yoshimura1988wall,jana1995apparent}. 

Confinement and slip become important when performing rheological measurements with suspensions of particles with diameters of 140 or \SI{250}{\micro\meter}. Slip is avoided using a 25\,mm-wide plate-plate crosshatched geometry with a typical roughness of \SI{1}{\milli\meter}. Confinement is a  trickier issue since increasing the gap thickness leads to a larger meniscus and thus to a significant error in viscosity measurements \citep{cardinaels2019quantifying}. 
We circumvent this issue by using a wide reservoir (a 5~cm-wide cup) mounted on the lower plate and filled with a 5 mm-thick layer of suspension ($\gtrsim 20d_p$) on which the upper tool is lowered to touch the free interface \citep{chateau2018pinch,chateau2019breakup}.
The additional torque exerted by the exceeding fluid in the reservoir can be computed analytically if the reservoir width and the suspension thickness are known \citep{vrentas1991exact}. We have implemented numerically this correction and we have been able to reproduce both published data and experimental results with and without the reservoir for Newtonian fluids and dense suspensions of 80-\si{\micro\meter} particles.

The relative bulk viscosity of monomodal granular suspensions, $\eta_s$, is defined as the viscosity of the whole suspension relative to that of the viscosity of the suspending fluid. It is independent of the particle size and shear rate and is solely an increasing function of the particle volume fraction $\phi$ that diverges at a maximum value, $\phi_c$, for which the suspension ceases to flow as shown in figure~\ref{fig:rheology}$(a)$. Note that the value $\phi_c\simeq 0.53$ inferred from curve fitting is provided for the plot but a more exhaustive study of the viscosity in the dense regime would be required to confirm this value. The value of $\phi_c$ depends on the frictional contacts between particles \citep[see e.g.][]{tapia2019influence}. It is important to note for the following discussion that $\phi_c$  also depends on the particle-size distribution for polydisperse suspensions.

For bimodal suspensions,  the viscosity not only depends on $\phi$ but also on the particle sizes, $d_1$ and $d_2$ ($d_1<d_2$), of the two species and on the fraction of small particles in the solid phase, $\zeta_{small}$, see figure~\ref{fig:rheology}$(b)$. Experimental studies of dense bimodal systems in the literature indicate that their viscosity is also controlled by the value of the maximum packing fraction, $\phi_c$, which is found to be higher than that of monomodal suspensions \citep{chang1994effect,chong1971rheology}. This increase in $\phi_c$ results from the ability of small particles to fill the holes between the large ones \citep{macosko1994rheology}. For bimodal suspensions with a given size ratio $d_2/d_1$, the relative bulk viscosity, $\eta_s$, is equal to that obtained in the sole presence of the large particles, when $\zeta_{small}=0\,\%$.  Increasing $\zeta_{small}$ leads to a decrease of $\eta_s$, down to a minimum located between $\zeta_{small}=25$ and 50\%, and then to an increase up to the value obtained for a suspension of small particles, $\zeta_{small}=100\,\%$. The minimum in viscosity is more pronounced with increasing $d_2/d_1$ see figure~\ref{fig:rheology}$(b)$. 

\subsection{Experimental apparatus}

\begin{figure}
    \centering
    \includegraphics[width=0.8\linewidth]{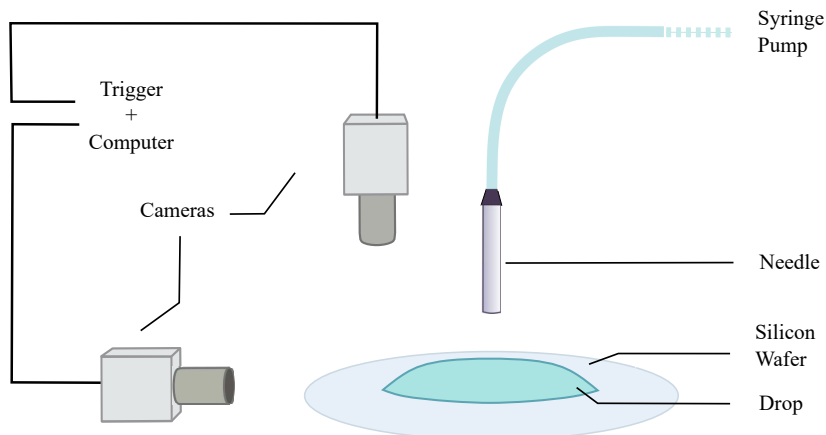}
    \caption{Sketch of the experimental apparatus}
    \label{fig:apparatus}
\end{figure}

We create droplets of suspension having a volume $V_0=$ \SI{300}{\micro\liter} (spherical radius $R_0=$ \SI{4.15}{\milli\meter})  with a syringe pump (flow rate \SI{10}{\milli\liter\per\min}), as shown in figure~\ref{fig:apparatus}. The droplets spread over a silicon wafer (Si-Mat) cleaned with ethanol and distilled water and dried with clean-room wipes (plasma cleaning did not change the results nor the quality of the data and was therefore considered an unnecessary precaution). 
Acquisition is made from side and top views with two synchronised monochrome cameras (Basler acA2440-35um, 2000$\times$2448 pixels) on which 1:1 macro objectives are mounted (I2S visions, MC series). High spatial resolution is required to capture the dynamics of the contact line. These optics give a resolution of \SI{3.45}{\micro\meter\per\pixel}. Frame rates from 5 to 10 fps are required for the side view, especially at the beginning as the drop spreads quickly. Frame rates of 0.5 fps are sufficient for the top view as most of  the interesting data are extracted at long times when spreading is slow and the slope of the drop interface is not too steep.
A typical set of experiments for a single suspension batch is made of 10 different runs of spreading drops. These runs are acquired after 10 unused runs (corresponding to a total amount of $\sim$ 3 ml) to avoid effects coming from the front of the advancing suspension in the tubing and in the needle. After performing these blank runs, the experiments are seen to be very reproducible at the desired volume fraction $\phi=0.4$.
Care is also taken to account for temperature and humidity variations. These two factors mainly impact the suspending fluid viscosity, $\eta_f$. To this end, systematic viscosity measurements are performed during the experiments using a capillary viscometer.

\subsection{Side-view analysis}

\begin{figure}
    \centering
    \includegraphics[width=0.9\linewidth]{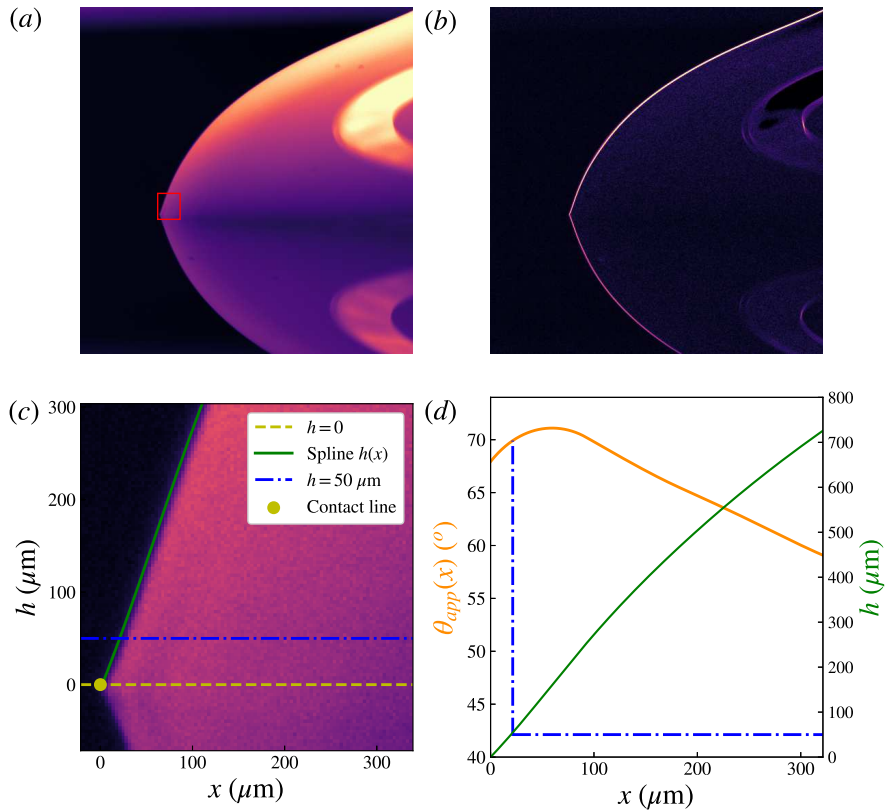}
    \caption{Data extraction from a picture of a spreading drop. Reflection on the wafer helps to detect the advancing contact line. $(a)$ Raw picture and $(b)$ Sobel filtering. The red rectangle in the top-left panel corresponds to the blown-up region in panel $(c)$, showing the fitted spline of the drop profile (green curve), the position of the contact line (yellow dot), and the drop height $h=$~\SI{50}{\micro\meter} (dash-dotted blue line). $(d)$ Results extracted from the fit: drop height, $h$, as a function of the distance to the contact line, $x$ (green), and contact angle computed from the spline derivation according to $\theta_{app}(x)=\tan^{-1}\left(\frac{{\rm d}h}{{\rm d}x}\right)$ (orange).}
    \label{fig:data}
\end{figure}

We characterise the dynamics of spreading by measuring the dynamic contact angle $\theta_{app}$ as a function of the dimensionless contact line velocity $U$. Angle measurement from the side views has been improved upon the previous work of \citet{zhao2020spreading} owing to a better optical resolution and the possibility to perform an automated local angle measurement.
We tailor the background lighting of our system with masks so that the drops captured by the side camera appear bright on a dark background, as seen on the top panel of figure~\ref{fig:data}$(a)$. Drop shape detection is performed with the Sobel filter of the scikit-image package in Python, shown in figure~\ref{fig:data}$(b)$, and further thresholded to extract the contact line coordinates and the drop profile, $h(x)$, with $x=0$ being the position of the contact line, see figure~\ref{fig:data}$(c)$. The position of the contact line is defined as the leftmost bright pixel after image processing, as presented in figure~\ref{fig:data}$(b),(c)$. The drop profile is then smoothed with a cubic spline, as shown in figure~\ref{fig:data}$(c),(d)$. The contact line velocity, $U$, is obtained by locating the triple contact line while the apparent dynamic contact angle, $\theta_{app}$, is inferred from the derivation of this spline. A spline is a piece-wise polynomial function, defined by the number of points, or knots, that it passes through. Spline functions are regular at zero, first, and second order of derivation, as demonstrated in figure~\ref{fig:data}$(d)$. To optimise spline adjustment to the data at all slopes, a small knot-to-knot distance is chosen at early times and then increased as the drop flattens. This process prevents meaningless oscillations at small slopes (large distance between knots). For all pictures, improper fits due to loss of focus or dust are discarded. The results are found to be similar to those obtained by adjusting manually a straight line to the air/liquid interface near the contact line in the range $ 30^{\circ} \lesssim \theta_{app}\lesssim 85^{\circ}$.  The fitted profile can then be derived once or twice at any point of the interface. The possibility of changing the measurement height, $h$, of the contact angle is one of the major benefits of this automated numerical procedure compared to previous manual measurements, see figure~\ref{fig:data}$(d)$.

\subsection{Top-view analysis}

Top views are used to visualise the structure of the particle network near the contact line and also to measure the distance between the particles and the contact line, $L$. Measurements using the ImageJ FiJi software package \citep{schindelin2012fiji} are performed over roughly 30 particles at the front. The precision of these measurements is set by the resolution of the pictures (3.45 \si{\micro\meter\per\pixel}).

\section{Identifying the region of validity of the Cox-Voinov law}
\label{sec:wetting}
\subsection{Simple fluids}

\begin{figure}
    \centering
     \includegraphics[width=0.9\linewidth]{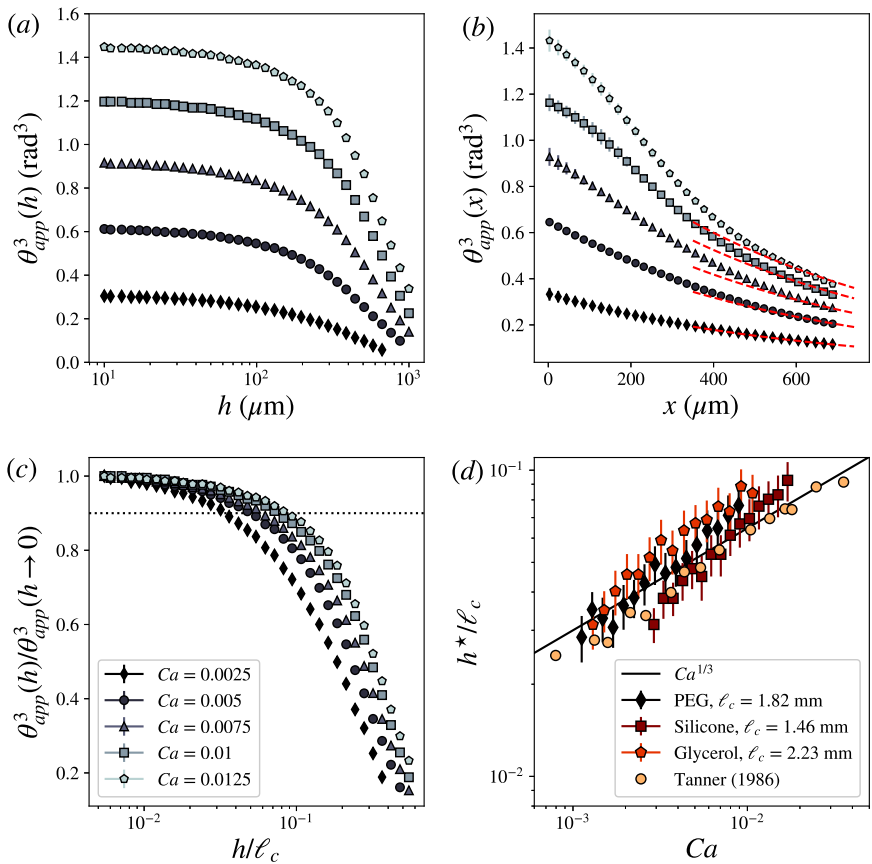}
    \caption{Cube of the contact angle, $\theta_{app}^3$, versus $(a)$ the measurement height, $h$, and $(b)$ the horizontal distance to the contact line, $x$, averaged over 7 experimental runs for 5 capillary numbers, $Ca = 0.0025\,(\lozenge),0.005\,(\circ),0.0075\,(\triangle),0.01\,(\square),0.0125\,(\pentagon)$, using a Newtonian fluid (PEG copolymer).
    Red dashed lines: static shape \eqref{eq:sol_static} for $R=8.5,\,8.7,\,8.9,\,9.3$, and \SI{10}{\milli\meter} (estimated as to provide the best fit with the experimental data) and $\ell_c=$~\SI{1.82}{\milli\meter}. $(c)$ Normalised cube of the contact angle, $\theta_{app}^3/\theta_{app}^3(h\rightarrow0)$, versus normalised height, $h/\ell_c$. Black dotted line: threshold for the plateau length at $\theta^3(h^\star)/\theta^3(h\rightarrow0)=0.9$ . $(d)$ Normalised transition height, $h^\star/\ell_c$, versus $Ca$ for 3 different Newtonian fluids: PEG (black $\lozenge$), Silicone oil V1000 (maroon $\square$) ($\rho=970$~kg/m$^3$, $\gamma=21$~mN/m, $\eta=1.0$~Pa.s, $\ell_c=1.46$~mm), and glycerol (red $\pentagon$) ($\rho=1260$~kg/m$^3$, $\gamma=63$~mN/m, $\eta=1.3$~Pa.s, $\ell_c=2.23$~mm) as well as the inflection-point measurements (gold $\circ$) of \cite{tanner1986edge}  with highly viscous Silicone oil. Black solid line: $h^\star/\ell_c=0.3\,Ca  ^{1/3}$.}
    \label{fig:height_PEG}
\end{figure}

In \S\,\ref{subsec:intermediate}, we have identified the length scale, $h^\star$, that characterises the height of the interface at the transition between the viscous-capillary regime governed by the Cox-Voinov relation \eqref{eq:voinovTannerLaw} and the viscous-capillary-gravity regime where gravity starts to prevail and where an inflection point should exist. We start our experimental characterisation of $h^\star$ by investigating the shape of the interface in the case of simple fluids. 

Inspired by the form of the Cox-Voinov law \eqref{eq:voinovTannerLaw}, we measure, for capillary numbers in the range $0.0025\leq Ca\leq 0.0125$, how the cube of the angle between the interface and the horizontal, $\theta_{app}^3$, depends on the interface height, $h$, and on the horizontal distance to the contact line, $x$, see figure~\ref{fig:height_PEG}$(a),(b)$, respectively. The data come from the average over 7 experimental runs. We identify two regions. Starting from the contact line, for any capillary number, $\theta_{app}^3$ is first independent of the height at which it is measured, see figure~\ref{fig:height_PEG}$(a)$. A plateau can thus be defined for $h$ ranging from 10 to almost \SI{100}{\micro\meter}. Note that this plateau is better seen in semi-log scale in figure~\ref{fig:height_PEG}$(a)$. Discussion of these results in the light of the Cox-Voinov law would lead us to expect that $\theta_{app}^3$ is an increasing function of $h$. However, two experimental facts may prevent us from seeing this increase. First, an inflection point can be seen for some experiments, see for instance figure \ref{fig:data}$(d)$, but the process of averaging over several runs likely smears out the change of curvature. Second, the length scales that we can probe are at least three orders of magnitude larger than the nanometric cut-off scale, $\lambda$. We thus expect the logarithmic term to increase slowly with distance, leading to difficulties in distinguishing the shape of the interface from a straight line. Similar conclusions regarding the local slope near the contact line have been reached by \cite{rio2005gouttes}. It is worth mentioning that the plateau value increases with the capillary number as expected from the Cox-Voinov law \eqref{eq:voinovTannerLaw}. Away from the contact line, the angle decreases, suggesting a growing contribution of gravity to the force balance. In figure~\ref{fig:height_PEG}$(b)$, the experimental contact angle is plotted and the theoretical prediction obtained from the balance between gravity and capillarity is superimposed at largest distances from the contact line. We find that the agreement of  the static solution \eqref{eq:sol_static} with the data is satisfactory and improves with decreasing $Ca$, provided we leave the radius of the droplet as a fitting parameter since it cannot be inferred in our experiments. Indeed, the imaging field does not cover the whole drop as the focus is on the contact line region which requires a significant magnification.  

Comparison of the datasets is made easier if we normalise $\theta_{app}^3(h)$ by its asymptotic value $\theta_{app}^3(h\rightarrow0)$ and $h$ by the capillary length $\ell_c$, see figure\,\ref{fig:height_PEG}$(c)$. From these plots, we define the experimental transition height $h^\star$ as the height at which $\theta_{app}^3(h)$ departs from $\theta_{app}^3(h\rightarrow0)$ by 10\%. Figure~\ref{fig:height_PEG}$(d)$ shows the inferred $h^\star$ as a function of $Ca$ for different simple fluids. The normalised transition height, $h^\star/\ell_c$, increases as $Ca^{1/3}$ in agreement with the prediction \eqref{eq:vgcScale} of the dimensional analysis in \S\,\ref{subsec:intermediate}. We also report in this graph the measurements of the inflection point of the interface obtained by \cite{tanner1986edge}, although the present interpretation as an upper limit of dissipation was not mentioned in this work. These data agree well with the present estimates of $h^\star$ as well as with the $Ca^{1/3}$ scaling. Note that this scaling still holds when varying the threshold between 1\% and 15\%. The threshold of 10\% provides the best match with the measurements of \citet{tanner1986edge}. If we grant $h^\star$ the interpretation of a viscous-capillary cut-off length and refer to figure~\ref{fig:height_PEG}$(d)$, we can thus conclude that measuring contact angles at heights well below \SI{100}{\micro\meter} warrants probing the region of the droplet where the apparent dynamic contact angle is set by a balance between  viscous dissipation and capillarity only.

\subsection{Granular suspensions}
 
We now test the relevance of $h^\star$ to the spreading of drops of granular suspensions. Because they are density-matched to the suspending fluid and do not modify surface tension, the particles are expected to modify only viscous dissipation and to leave gravitational and capillary effects unchanged.

\begin{figure}
    \centering
    \includegraphics[width=0.9\linewidth]{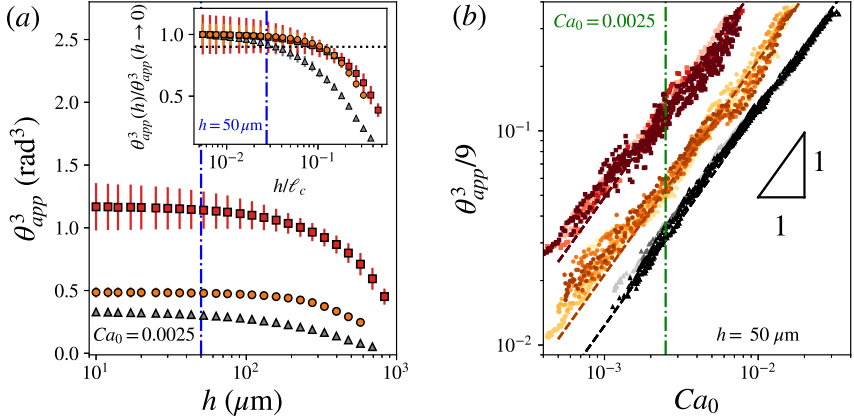}
\caption{(a) Cube of the contact angle, $\theta_{app}^3$, versus the measurement height, $h$, for different fluids: pure suspending fluid (grey $\triangle$), 10-\si{\micro\meter} monomodal suspension (red $\square$), and 10-\SI{80}{\micro\meter} bimodal suspension at $\zeta_{10}=50\,\%$ (orange $\circ$), resulting from the analysis of 7, 8, and 10 drop-spreading runs, respectively, and at the same fluid capillary number $Ca_0=\eta_f U/\gamma_f=0.0025$. Inset: normalised cube of the contact angle, $\theta_{app}^3/\theta_{app}^3(h\rightarrow0)$, versus normalised height, $h/\ell_c$. Blue dashed-dotted line: selected measurement height $h=$~\SI{50}{\micro\meter}. $(b)$ Variation of $\theta_{app}^3/9$ as a function of $Ca_0$ performed at a height $h=$~\SI{50}{\micro\meter} for the different runs for spreading drops made of pure suspending fluid (grey dots), 10~\SI{80}{\micro\meter} monomodal suspension (red dots), and 10-\SI{80}{\micro\meter} bimodal suspensions at $\zeta_{10}=50\,\%$ (orange dots). The different colour shades represent different experiments. The dashed lines correspond to the average of the linear fits of data coming from each run  and are used to infer the relative apparent viscosity, $\eta_w$.}
\label{fig:height_Suspension}
\end{figure}

Figure~\ref{fig:height_Suspension}$(a)$ presents the variation of $\theta_{app}^3$ with the measurement height, $h$, for monomodal granular suspensions of 10-\si{\micro\meter} particles and bimodal suspensions of 10-\SI{80}{\micro\meter} particles with $\zeta_{10}=50\,\%$, at a constant capillary number $Ca_0=\eta_f U/\gamma_f$, where subscript $0$ in the capillary number emphasises that it is computed using properties of the suspending fluid. Reference data for the pure suspending fluid at $Ca_0$ are also provided for comparison in figure~\ref{fig:height_Suspension}$(a)$. The behaviour of $\theta_{app}^3(h)$ for the suspensions is similar to that seen for the reference fluid, see also figure~\ref{fig:height_PEG}$(a,c)$. A plateau region is again observed close to the contact line, for $10\lesssim h\lesssim$~\SI{100}{\micro\meter}, while there is a decay at larger distances. The addition of the particles leads to an increase in the plateau value of $\theta_{app}^3$, which depends on the particle mixture components. Provided measurements are undertaken within the plateau region at constant height across all experiments, we can obtain an unequivocal apparent contact angle, at a given capillary number. The dependence between these two quantities is then \textit{a priori} interpretable in terms of the Cox-Voinov relation. In the following, the measurement height is set at $h=$~\SI{50}{\micro\meter}. This chosen height seems a good compromise between smaller heights having large measurement noise and larger heights that might fall outside of the region where viscosity matters. Note that taking $h$ in the range \SI{20}{\micro\meter} to \SI{100}{\micro\meter} yields similar results for the apparent wetting viscosity that will be introduced below.

Figure~\ref{fig:height_Suspension}$(b)$ displays typical variations of $\theta_{app}^3/9$ as a function of the capillary number of the suspending fluid, $Ca_0$. Data from different experimental runs are plotted for the same fluids as those used in figure~\ref{fig:height_Suspension}$(a)$. For a given fluid (suspensions or reference fluid), the tight collapse of the different $\theta_{app}^3(Ca_0)$ curves coming from the different runs bears witness to the good reproducibility of the experiments. All the datasets collapse on straight lines with unity slopes in log-log representation, i.e. $\theta_{app}^3/9\propto Ca_0$. For each run, a linear fit is performed and the average (dashed) line for a given fluid is the average of the corresponding linear fits. The measured angles typically lie between \SI{30}{\degree} and \SI{90}{\degree}, the upper limit being set by the algorithm. Tracking is interrupted below \SI{30}{\degree} due to the failure of the Cox-Voinov. Protrusions of the particles from the free surface are observed at small contact angles and might be responsible for this discrepancy.

As expected, the data from the suspending fluid are in excellent agreement with the Cox-Voinov law \eqref{eq:voinovTannerLaw}. The logarithm factor in \eqref{eq:voinovTannerLaw}, is found to be $\approx 11.9$ in good agreement with values in the literature \citep{voinov1976hydrodynamics}. It is important to note that the measurements are undertaken at a constant height,  $h$, within the plateau region and not at a constant $x$. However, since $x$ and $h$ are of the same order of magnitude and sufficiently large compared to $\lambda$, the logarithm factor is not varying significantly and can be considered constant.

For suspensions, data for $\theta_{app}^3(Ca_0)$ in figure~\ref{fig:height_Suspension}$(b)$ equally collapse on a straight line with a slope of unity in the log-log representation (red and orange dashed lines): the spreading of these materials still seems to follow a Cox-Voinov relation \eqref{eq:voinovTannerLaw}. However, since the values of $\theta_{app}^3$ in the plateau are changed by the addition of particles as shown in figure~\ref{fig:height_Suspension}$(a)$, the different $\theta_{app}^3(Ca_0)$ lines are shifted with respect to each other in figure~\ref{fig:height_Suspension}$(b)$, in agreement with previous experiments \citep{zhao2020spreading}. Assuming that the logarithmic factor has the same order of magnitude for the pure suspending fluid and for suspensions (as we do not expect the particles to modify the mechanics at the nanometric scale), we can superimpose all curves and recover the Cox-Voinov law \eqref{eq:voinovTannerLaw} for all liquids if we adjust the viscosity used in the capillary number, $Ca=\eta U/\gamma$, with $\eta=\eta_f\eta_w$ where $\eta_w$ is the relative apparent wetting viscosity of the suspensions.  Again, the shift and consequently the apparent wetting viscosity strongly depends on the particle mixture components. Here, we see that the suspension of 10-\si{\micro\meter} particles has a larger apparent wetting viscosity than the 10-80-\si{\micro\meter} bimodal suspension. The behaviour of the apparent viscosity, $\eta_w$, is examined in detail in the following \S\,\ref{sec:susp_wetting} for suspensions consisting of different particle combinations.

\section{Probing dissipation with particles}
\label{sec:susp_wetting}
%

\subsection{Suspensions with particles having a large difference in size}
\label{subsec:10-80}

\begin{figure}
    \centering
  $\vcenter{\hbox{\includegraphics[width=0.9\linewidth]{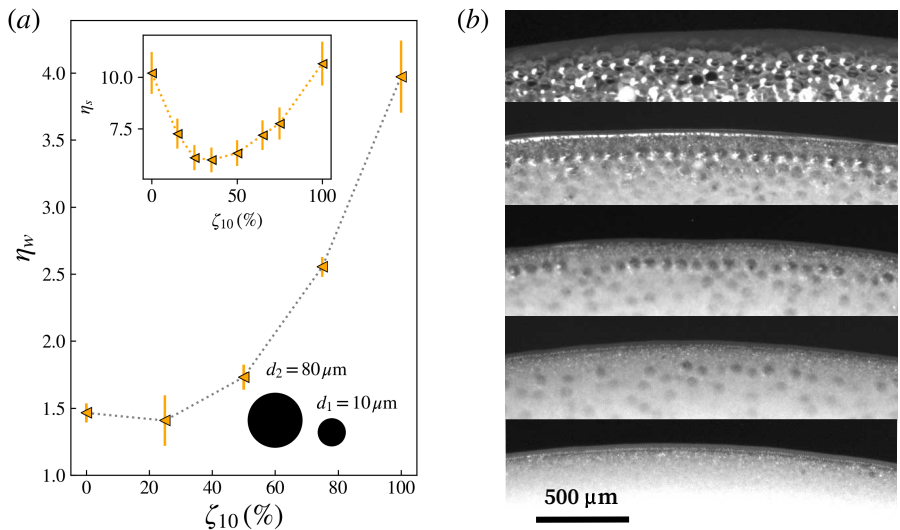}}}$
  \caption{Experimental results for suspension blends with fixed $d_1=$~\SI{10}{\micro\meter} and $d_2=$~\SI{80}{\micro\meter} and varying $\zeta_{10}$ at $\phi=0.4$. $(a)$ Relative wetting viscosity, $\eta_w$, as a function of the fraction of the small particles in the solid phase, $\zeta_{10}$.  Inset: bulk viscosity, $\eta_s$, of the corresponding suspensions versus $\zeta_{10}$. The dotted lines are guides for the eyes. $(b)$ Top-view pictures taken when $\theta_{app}=35^{\circ}$ at $h=$~\SI{50}{\micro\meter} for $\zeta_{10}=0,\,25,\,50,\,75,\,100\,\%$ from top to bottom, respectively.}
    \label{fig:10-80microns}
\end{figure}

We focus first on suspensions consisting of particles having a large difference in diameter, $d_1=$~\SI{10}{\micro\meter} and $d_2=$~\SI{80}{\micro\meter} ($d_2/d_1=8$). Figure~\ref{fig:10-80microns}$(a)$ shows that the relative wetting viscosity of these suspensions, $\eta_w$, increases with increasing fraction of the small particles, $\zeta_{10}$, with a  steeper growth beyond $\zeta_{10} \approx 50\,\%$. This behaviour is in stark contrast with the $\zeta_{10}$-dependence  of the relative bulk viscosity of the same mixtures shown in figure~\ref{fig:rheology}$(b)$ and plotted again in the inset in figure~\ref{fig:10-80microns}$(a)$.  The relative bulk viscosity of the blends presents a minimum at $\zeta_{10}=35\,\%$, i.e. a mixture containing roughly a third of small particles. This minimum viscosity is 40\,\% smaller than the values obtained in the monomodal cases at $\zeta_{10}=0\,\%$ and 100\,\% (consisting of monomodal suspensions of 80 and \SI{10}{\micro\meter}, respectively). It is worth noting that, even in the monomodal case with the small particles, $\eta_w$ is smaller (by a factor of order 2) than the corresponding value of its bulk viscosity, $\eta_s$. This inability to reach the bulk value is always observed even for the smallest particles as there is always a small region devoid of particles at the tip followed by an ordered region, as discussed in the following and in \S\,\ref{sec:conclusion}.

Top-view pictures of the suspension near the moving contact line provide information about the origin of the discrepancy between the apparent viscosity extracted from bulk rheology and its counterpart extracted from the dynamics of spreading, see figure~\ref{fig:10-80microns}$(b)$. As previously shown in the monomodal case \citep{zhao2020spreading}, there is a pure-fluid region devoid of particles near the contact line. We observe a similar region in the monomodal and bimodal cases depicted in figure \ref{fig:10-80microns}$(b)$. The extent of this region is larger in the case of the 80-\si{\micro\metre} particles, $\zeta_{10}=0$\% (top picture), than in the 10-\si{\micro\metre} case, $\zeta_{10}=100$\% (bottom picture). For intermediate values of $\zeta_{10}$, the 10-\si{\micro\metre} particles are able to move in-between the 80-\si{\micro\metre} particles. The particles closest to the contact line arrange in an orderly, crystalline structure. This ordering is particularly marked for the 80-\si{\micro\meter} particles, seen on the top pictures of figure \ref{fig:10-80microns}$(b)$. The large 80-\si{\micro\meter} particles seem to be maintained at the same distance from the contact line as small 10-\si{\micro\meter} particles are added, i.e. as $\zeta_{10}$ is increased, but their linear density decreases.

The images displayed in figure \ref{fig:10-80microns}$(b)$ are obtained for the same dynamic contact angle, $\theta_{app}=35$~\si{\degree}. The real capillary number of all these experiments (using the effective viscosity of the suspension, $Ca = \eta_wCa_0$) is thus constant, and equal here to approximately $2\cdot10^{-3}$, leading to $h^\star\sim 100$ \si{\micro\meter} according to \eqref{eq:vgcScale} and using the prefactor inferred from the figure~\ref{fig:height_PEG}$(d)$, i.e.\, $h^\star\approx 0.3\,\ell_c\,Ca^{1/3}$. The discrepancy between the viscosity measured in the bulk and that estimated from spreading experiments is now clear. The 80-\si{\micro\meter} particles experience strong confinement in the viscous-capillary region since $h^\star\sim d_{2}$, and only a few of them can penetrate this region. In contrast, $h^\star$ is 10 times larger than the diameter of the 10-\si{\micro\metre} particles. Even if these small particles experience confinement close to the contact line, a significant part of the viscous-capillary region is filled with a dense phase of small particles akin to a suspension bulk. Thus, the contribution of the 10-\si{\micro\metre} particles to dissipation in the Cox-Voinov region is expected to be larger than that of the 80-\si{\micro\metre} particles, as reported in experiments with monomodal suspensions. In the case of the bimodal suspensions studied in this section, increasing the small particle fraction, $\zeta_{10}$, is expected to lead to a continuous increase in $\eta_w$, as observed in figure~\ref{fig:10-80microns}$(a)$.

\subsection{Varying the size of the large particles}
\label{subsec:varyingd2}

\begin{figure}
    \centering
 \includegraphics[width=0.9\linewidth]{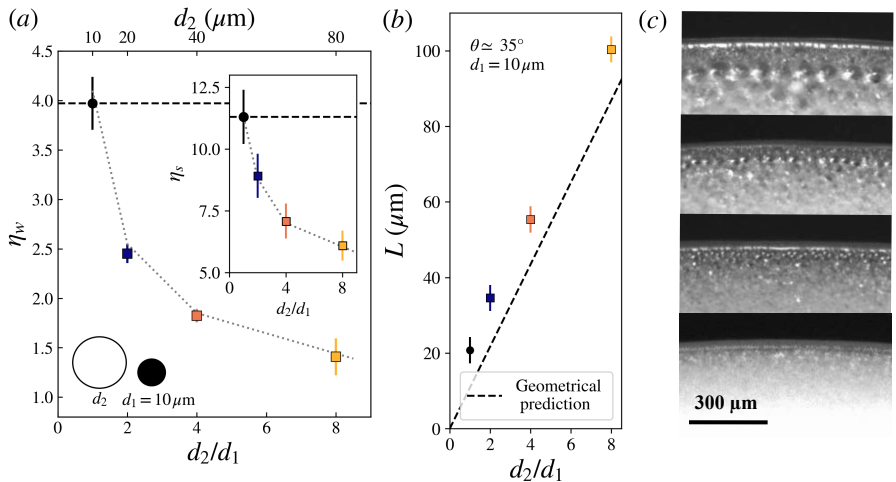}
    \caption{Experimental results for \SI{10}{\micro\meter} monomodal suspension ($\bullet$ symbols) and bimodal blends with $d_1=$~\SI{10}{\micro\meter} and varying $d_2$ (= 20, 40, \SI{80}{\micro\meter}) at fixed $\zeta_{10}=25\,\%$ and $\phi=0.4$ ($\square$ symbols). $(a)$ Relative wetting viscosity, $\eta_w$, as a function of the size ratio, $d_2/d_1$. The horizontal dashed line indicates the viscosity value for the monomodal suspension consisting of the sole small particles of size \SI{10}{\micro\meter}. Inset: bulk viscosity, $\eta_s$, of the corresponding suspensions versus $d_2/d_1$. The grey dotted lines are guides for the eyes.  $(b)$~Distance of approach of the large particles for the bimodal suspensions with $d_1=$~\SI{10}{\micro\meter} and $\zeta_{10}=25\,\%$ ($\square$ symbols) and the \SI{10}{\micro\meter} monomodal suspension ($\bullet$). The dashed line corresponds to the geometrical prediction \eqref{eq:distance_CL}. $(c)$ Top-view pictures taken when $\theta_{app}=35^{\circ}$ at $h=$~\SI{50}{\micro\meter} for monomodal suspension of size \SI{10}{\micro\meter} and bimodal suspensions with $(d_1,\zeta_{10})=($\SI{10}{\micro\meter}$,25\,\%)$ and $d_2$ = 20, 40, \SI{80}{\micro\meter} (from bottom to top, respectively).}
    \label{fig:wetting_viscosity10}
\end{figure}   

We now turn to bimodal suspensions consisting of 10-\si{\micro\meter} particles but with large particles of variable diameter (i.e. $d_2= 20, 40,$ \SI{80}{\micro\meter}) at $\zeta_{small}=25\,\%$. Bimodal suspensions formulated this way should have a bulk viscosity close to the minimum observed on the curves of figure~\ref{fig:rheology}$(b)$. Figure~\ref{fig:wetting_viscosity10}$(a)$ shows that the apparent wetting viscosity of these suspensions is maximum in the monomodal case and decreases for bimodal suspensions as the large particle size increases, i.e. with increasing $d_2/d_1$. This result again confirms that particles with a diameter much smaller than $h^\star$ increase the dissipation in the Cox-Voinov region. This trend is similar to that of the bulk viscosity, $\eta_s$, with increasing $d_2/d_1$ for $\zeta_{small}=25\,\%$, see the inset figure~\ref{fig:wetting_viscosity10}$(a)$. However, the magnitude of the size-ratio effect comes from two distinct physical origins. For example,  the value of the apparent wetting viscosity, $\eta_w$, for $d_2/d_1=8$ is close to that of the suspending fluid whereas the value of the bulk viscosity, $\eta_s$, is much larger.  The decrease of the bulk viscosity comes from the size ratio affecting the maximum packing fraction $\phi_c$ (larger for large $d_2/d_1$), while the wetting viscosity reduction comes from confinement effects near the contact line as the dissipation is greater for particles which can efficiently reach the dissipation region near the contact line ($d\ll h^\star$). In the case of figure~\ref{fig:wetting_viscosity10}, the small particles are participating less in the dissipation as $\zeta_{10}$ is small and the overall dissipation is mostly set by the large particles.

Figure \ref{fig:wetting_viscosity10}(b) shows that the distance of approach of the large particles, $L$, increases with their diameter. This measured length follows the prediction of a geometrical model describing the minimal distance of approach of a particle in a wedge,
\begin{equation}
    L=\frac{d}{2}\left[\frac{1}{\sin(\theta_{app})}+\frac{1}{\tan(\theta_{app})}-1\right],
    \label{eq:distance_CL}
\end{equation}  
where $\theta_{app}$ is the apparent contact angle near the contact line. We find that this distance is not sensibly affected by the presence of the small particles. The relation \eqref{eq:distance_CL} holds both in monomodal and bimodal using the diameter of the corresponding particles. In the monomodal case, it determines the size of the depleted region \citep{zhao2020spreading}.

The top-view pictures in figure~\,\ref{fig:wetting_viscosity10}$(c)$ confirm the existence of size segregation at the contact line as already mentioned in \S\,\ref{subsec:10-80}. The large particles form ordered rows behind the small particle region as already noted in \S\,\ref{subsec:10-80}. However, we observe that the 10-\si{\micro\meter} particles do not go through the network formed by the 20-\si{\micro\meter} particles, see bottom image of figure \ref{fig:wetting_viscosity10}(c). We move to this aspect of suspension spreading in the following section with clearer visualisations with larger particles.

\subsection{Varying the size of the small particles}

\begin{figure}
    \centering
    \includegraphics[width=0.9\linewidth]{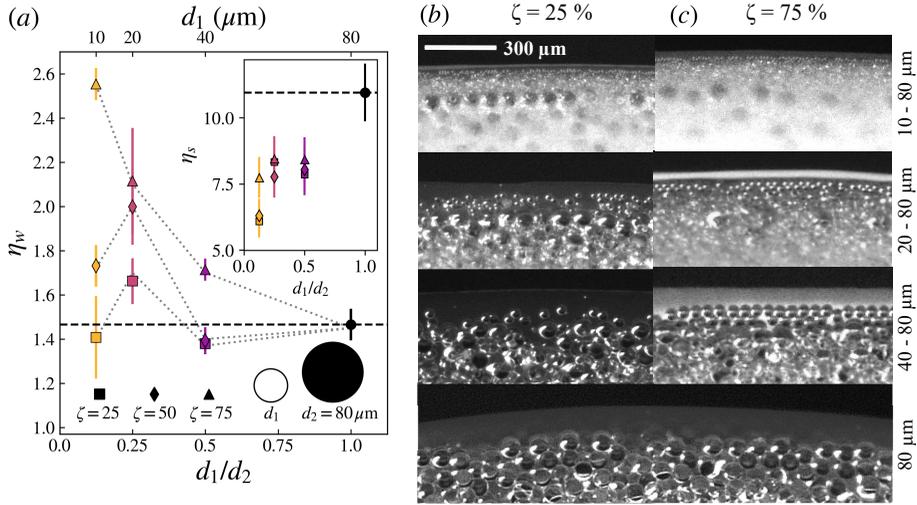}
    \caption{Experimental results for suspension blends with $d_2=$~\SI{80}{\micro\meter} and varying $d_1$ (= 10, 20, \SI{40}{\micro\meter}) at $\zeta_{small}=25\,\%$ ($\square$ symbols), $\zeta_{small}=50\,\%$ ($\lozenge$ symbols) and $\zeta_{small}=75\,\%$ ($\triangle$ symbols). $(a)$ Relative wetting viscosity, $\eta_w$, as a function of $d_1/d_2$. The dashed line indicates the viscosity value for the monomodal suspension consisting of the sole large particles of size \SI{80}{\micro\meter}. Inset: bulk viscosity of the corresponding suspensions. $(b)$ and $(c)$ Top-view pictures taken when $\theta_{app}=35^{\circ}$ at $h=$~\SI{50}{\micro\meter} for bimodal suspensions with $d_2=$~\SI{80}{\micro\meter}, $\zeta_{small}=25\,\%$ $(b)$ and $\zeta_{small}=75\,\%$ $(c)$. Small particle size is increasing from top to bottom ($d_1=$10, 20, \SI{40}{\micro\meter} and monomodal suspension of size \SI{80}{\micro\meter}).}
    \label{fig:wetting_viscosity80}
\end{figure}

Figure~\ref{fig:wetting_viscosity80}$(a)$ gathers the wetting viscosity, $\eta_w$, of the monomodal \SI{80}{\micro\meter} suspension and bimodal blends having a fixed $d_2=$~\SI{80}{\micro\meter} and varying $d_1$ and $\zeta_{small}$. When the solid phase mainly consists of small particles ($\zeta_{small}=75\,\%$), $\eta_w$ decreases with increasing $d_1$ as seen in figure~\ref{fig:wetting_viscosity80}$(a)$. This observation confirms again that the wetting viscosity is set by the possibility for particles having a diameter much smaller than $h^\star$ to approach close to the contact line. However, a continuous decrease of $\eta_w$ with $d_1$ is absent when $\zeta_{small}=25\,\%$ or $\zeta_{small}=50\,\%$, i.e. when large particles constitute a significant portion of the solid blend. The relative wetting viscosity, $\eta_w$, reaches a maximum around $d_1/d_2=0.25$ but it otherwise shows low values, even lower than those obtained for the monomodal suspensions of 80-\si{\micro\meter} particles for $d_1/d_2 = 0.5$. In other words, for $\zeta_{small}=25-50\,\%$, the apparent wetting viscosity of the 10-\SI{80}{\micro\meter} and 40-\SI{80}{\micro\meter} blends are particularly low compared to that of the 20-\SI{80}{\micro\meter} and even to that of the monomodal 80-\si{\micro\meter} suspension.

The top views shown in figure~\ref{fig:wetting_viscosity80}$(b)$ provide some clues about the variation of the apparent wetting viscosity. At $\zeta_{small}=25\,\%$, for small values of $d_1/d_2$ (two topmost pictures), small particles flow through the large particle network and get close to the contact line. In contrast, for $d_1/d_2=0.5$ (pictures on the third row from the top), the presence of 40-\si{\micro\meter} particles disrupts the ordering of the 80-\si{\micro\meter} particles, compared to the other cases. A similar trend is observed for $\zeta_{small}=50\,\%$. In contrast, for $\zeta_{small}=75\,\%$, there is a large amount of ordered small particles near the contact line in front of the large spheres, as seen in figure~\ref{fig:wetting_viscosity80}$(c)$.

\begin{figure}
    \centering
      \includegraphics[width=0.4\linewidth]{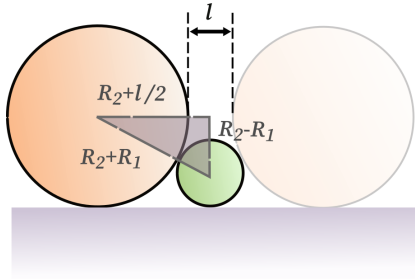}
        \caption{Sketch of two large particles separated by a small one on a solid plane.}
    \label{fig:matrix}
\end{figure}

Geometrical considerations can explain the effect observed for $d_1/d_2=0.5$. Two large particles with a radius $R_2=d_2/2$ sitting on the surface of a solid come to contact if the distance, $l$, between them goes to zero, see figure \ref{fig:matrix}. A small particle of radius $R_1$ can pass through the hole between the two particles and the solid surface if its radius is at most $ R_1=R_2/4$. Consequently, while the 10- and 20-\si{\micro\meter} particles can flow through the interstices created by the large \SI{80}{\micro\meter} particles, the 40-\si{\micro\meter} particles cannot. Instead, these particles induce defects in the 80-\si{\micro\meter} particle network, leading to the distortion of the large-particle matrix when the portion of small particles is not too large ($\zeta_{small} \lesssim 50\,\%$). This loss of organisation seems detrimental to the local dissipation. These geometrical considerations coupled with the wetting viscosity measurements suggest that the structure and therefore the local solid volume fraction near the contact line is a key element to explain the value of the wetting viscosity of granular suspensions.

However, the above geometrical arguments cannot explain the lower values of $\eta_w$ for the 10-\SI{80}{\micro\meter} blends compared to those of the 20-\SI{80}{\micro\meter} blends for $\zeta_{small}=25-50\,\%$ in figure~\ref{fig:wetting_viscosity80}$(a)$. Instead, this lessening effect may be attributed to bulk effects impacting on the variation of the wetting viscosity. The inset of figure~\ref{fig:wetting_viscosity80}$(a)$ indicates that the bulk viscosity is significantly smaller for the 10-\SI{80}{\micro\meter} blends than for the 20-\SI{80}{\micro\meter} blends for a given value of $\zeta_{small}$. Moreover, the bulk viscosity for $\zeta_{small}=25-50\,\%$ is lower than for $\zeta_{small}=75\,\%$ as it corresponds to the minimum bulk viscosity for bimodal blends, as seen in figure~\ref{fig:rheology}$(b)$. Bulk effects are therefore evidenced in this specific example and impact the value of $\eta_w$ as they happen to overcome size effects.

It is worth noting that, in the monomodal case, variation of the wetting viscosity can only come from confinement effect near the contact line as the bulk viscosity does not vary with particle size, as shown in figure~\ref{fig:rheology}$(a)$. The case of fixed $d_2$ and increasing $d_1$ evidences two competing effects: confinement effect governing the particle ability to approach the contact line and the variation of the bulk viscosity with $\zeta_{small}$ and $d_1/d_2$. In conclusion, for bimodal suspensions, a high wetting viscosity results from a complex compromise between a high fraction of small particles, $\zeta_{small}$, a large size ratio, $d_2/d_1$, and a high bulk viscosity, $\eta_s$, in addition to size effects already demonstrated.

\section{Concluding remarks}
\label{sec:conclusion}
The spreading of large drops onto a solid substrate is a rich and complex phenomenon by itself. Adding particles affects the spreading dynamics in a non-trivial way as there exists a high degree of inhomogeneity in the particulate drop going from a random solid bulk phase far from the contact line ($h\gg d$), to ordered dense monolayers when particles undergo confinement ($h\gtrsim d$), and finally to a particle-depleted region in the very close vicinity of the contact line ($h<d$). Using granular suspensions nonetheless offers an interesting way to investigate dynamical wetting as one can take advantage of this complex particulate organisation to modify locally the dissipation near the contact line. 

Dimensional analysis of the equation governing the spreading of large drops ($Bo\gtrsim1$) shows that the profile of the liquid/gas interface between the droplet and the atmosphere is set by a viscous-capillary balance as long as its height is smaller than a length scale $h^\star\sim\ell_c Ca^{1/3}$ beyond which the drop profile becomes also sensitive to gravity. This scaling has been validated in the present work for both simple fluids and suspensions and care was taken to perform measurements below this length scale (typically  \SI{50}{\micro\meter}) to ensure that gravity is negligible across the range of capillary numbers that we can probe. Drops of granular suspensions were then seen to follow the Cox-Voinov law relating the contact angle to the capillary number in a similar way as found for regular Newtonian fluids. However, the relative apparent viscosity, $\eta_w$, involved in the capillary number was found to differ from that of the bulk suspension, $\eta_s$.

In the present experiments, the typical magnitude of $h^\star\sim$ \SI{100}{\micro\meter} lies in the size range of the non-Brownian particles used. Therefore, dissipation is affected by particles during the spreading when $d\lesssim h^\star$ while the spreading dynamics is close to that of the pure fluid when $d\gtrsim h^\star$, i.e.\,when particles can not reach the viscous-capillary region. These geometrical considerations help rationalise that the relative wetting viscosity, $\eta_w$, depends on the particle size, $d$, unlike the relative bulk viscosity, $\eta_s$, which only depends on particle volume fraction, $\phi$. The relative wetting viscosity, $\eta_w$, is found to be maximum for the smallest particles and decreases to values close to the suspending fluid viscosity for $d\gtrsim$~\SI{100}{\micro\meter} even though $\eta_s$ is much larger than $\eta_w\simeq\eta_f$ for these dense suspensions ($\phi=0.4$), see figure~4 of \cite{zhao2020spreading}. This size cut-off of the wetting viscosity is thus roughly of the order of $h^\star$. This confirms that particles affect the wetting dynamics only if their size is small enough for them to reach the region dominated by viscosity. 

It may seem surprising that the Cox-Voinov law applies to such a complex system. Indeed, the possibility for the particles to occupy the viscous-capillary region in the case of a spreading droplet  evolves with time as $h^\star\propto Ca^{1/3}\propto U^{1/3}$ decreases with increasing time. However, in the present experiments, the variation of $Ca$ over only one decade prevents a strong change in $h^\star$. Performing measurements in a much lower $Ca$-range, while rather challenging, may reveal intriguing effects.

Predicting the value of the apparent wetting viscosity when $d\lesssim h^\star$ remains a difficult task. The value of the bulk viscosity, $\eta_s$, certainly affects the wetting viscosity, $\eta_w$. However, knowing whether this observation comes from long-range effects of the bulk phase far from the contact line is not easy. One must also keep in mind that $\eta_s$ varies with $\phi_c-\phi$ and strongly diverges as $\phi \rightarrow \phi_c$. For bimodal suspensions, a lower $\eta_s$ is the signature of a higher $\phi_c$ for the corresponding solid phase and therefore produces a lower $\eta_w$, as seen in the present experiments.

Another fundamental ingredient to consider in predicting $\eta_w$ is the possibility for the particles to crystallise in the viscous region. For monomodal suspensions, a monolayer of several rows of ordered particles is always observed in the vicinity of the contact line. In the bimodal case, crystallisation of the two population sizes can be hindered when the small particles cannot flow through large-particle holes, leading to a significant diminution of the apparent wetting viscosity. The structure of the particulate network near the contact line therefore directly impacts the drop spreading dynamics captured in $\eta_w$. It also explains that even for the smallest particles for which $d<h^\star$, the wetting viscosity, $\eta_w$, cannot reach the bulk value, $\eta_s$, because of this ordered region which possesses a lower viscosity.

This ordering very likely comes from the confinement of the particles near the contact line combined with the pressure of the dense particulate phase farther out. Rheological measurements combined with direct visualisation of dense colloidal suspensions ($\phi=0.52$) have clearly established the relation between a significant viscosity drop and moderate confinement, owing to the ordering of the particulate phase in sliding layers \citep{ramaswamy2017}. Conversely, for much dilute granular suspensions ($\phi\leq 0.2$), rheological measurements have exhibited a monotonous increase in dissipation for gap sizes of ten diameters and less \citep{peyla2011new}. This apparent disagreement could be explained by the nature of the solid phase (colloidal/granular) or the dissipation mechanisms  (friction/hydrodynamics) that depends on the solid volume fraction \citep{gallier2014rheology}. Indeed, at low volume fraction, dissipation is mainly due to hydrodynamic interactions while frictional contacts dominate for dense suspensions and may crucially depend on the local solid structure \citep{guazzelli2018rheology}.

With confinement strengthening, complex behaviours become even more apparent, e.g.\,viscosity minima when the gap thickness is commensurate with the particle diameters \citep{ramaswamy2017}. Such oscillatory values of the viscosity as a function of the gap have been reported under high confinement and high shear in numerical simulations of dense granular suspensions  \citep{fornari2016rheology}. In the present experiments, confinement effects may however be less easy to quantify for the following reasons. Confinement is changing over time as the drop spreads and the local fluid thickness varies with the radial position. Moreover, it is set by a solid surface and a free interface that can deform to relax high stress, contrary to solid boundaries in a rheometer and most of the simulations.

Finally, the present work may help understand the spreading of suspensions that are closer to those used in industrial processes and have a wide range of particle sizes \citep[e.g.\,from sub-micron to hundreds of microns in cement paste, see][]{bentz1999effects,celik2009effects}. Bimodal suspensions can be seen as a first step toward polydisperse solid blends as a smart choice in particle size and fraction in a bidisperse system can partially mimic an equivalent polydisperse suspension \citep{pednekar2018bidisperse}.

Declaration of Interests. The authors report no conflict of interest.
\bibliographystyle{jfm}
\bibliography{Pelosse_JFM_2022}
%
\end{document}